\documentclass[useAMS,usenatbib]{mn2e}

\usepackage{graphicx}
\usepackage{multirow}
\usepackage{array}

\def\width9{{\sc WIDTH9}}

\def\logg{\log(g)}
\def\teff{T_{\rm eff}}

\def\kms{km/s}
\def\bs{\langle B_{\mathrm{s}} \rangle}

\def\halpha{\mathrm{H}\alpha}

\def\synth3{{\sc SYNTH3}}

\def\synthmag{{\sc Synthmag}}
\def\synmast{{\sc Synmast}}

\def\vsini{\upsilon\sin i}

\def\b{|\mathrm{\mathbf{B}}|}

\def\gammaw{\gamma_\mathrm{Waals}}
\def\etal{et al.}

\def\ll{\lambda\lambda}

\def\ca{\textit{A}}
\def\cb{\textit{B}}
\def\cc{\textit{C}}
\def\fgammaw{f(\gammaw)}
\newcommand{\abn}[1]{\alpha(\mathrm{#1})}

\def\aj{AJ}%
\def\apj{ApJ}%
\def\apjs{ApJS}%
\def\aap{A\&A}%
\def\mnras{MNRAS}%
\newcommand{\inst}[1]{$^{#1}$}

\newcolumntype{+}{>{\global\let\currentrowstyle\relax}}
\newcolumntype{^}{>{\currentrowstyle}}

\title[Rotation, magnetism, and metallicity of M dwarf systems]{Rotation, magnetism, and metallicity of M dwarf systems
\thanks{Based on observations made with ESO Telescopes at
    the Paranal Observatories under programme ID 81.D-0189}}
\author[D. Shulyak \etal]{
D. Shulyak\inst{1}\thanks{E-mail: denis.shulyak@gmail.com}, A. Seifahrt\inst{1,2}, A. Reiners\inst{1}, O. Kochukhov\inst{3}, N. Piskunov\inst{3}\\
\inst{1}Institute of Astrophysics, Georg-August-University, Friedrich-Hund-Platz 1, D-37077 G\"ottingen, Germany\\
\inst{2}Department of Physics, University of California, One Shields Avenue, Davis, CA 95616, USA\\
\inst{3}Department of Physics and Astronomy, Uppsala University, Box 516, 751 20, Uppsala, Sweden}

\begin{document}

\date{\today}

\pagerange{\pageref{firstpage}--\pageref{lastpage}} \pubyear{2011}

\maketitle

\label{firstpage}

\begin{abstract}
{Close M-dwarf binaries and higher multiples allow the
  investigation of rotational evolution and mean magnetic flux
  unbiased from scatter in inclination angle and age since the
  orientation of the spin axis of the components is most likely
  parallel and the individual systems are coeval. 
  Systems composed of an early (M0.0~--~M4.0) and a late
  (M4.0~–-~M8.0) type component offer the possibility to study
  differences in rotation and magnetism between partially and fully
  convective stars.  }
{We have selected 10 of the closest dM systems to determine
  the rotation velocities and the mean magnetic field strengths based
  on spectroscopic analysis of FeH lines of Wing-Ford transitions at
  $1$~$\mu$m observed with VLT/CRIRES. We also studied the
  quality of our spectroscopic model regarding atmospheric parameters
  including metallicity. }
{ A modified version of the Molecular Zeeman Library (MZL) was used to
  compute Land\'e g-factors for FeH lines.  Magnetic spectral synthesis
  was performed with the \synmast\ code.  }
{ We confirmed previously reported findings that less massive M-dwarfs
  are braked less effectively than objects of earlier types. Strong
  surface magnetic fields were detected in primaries of four systems
  (GJ~852, GJ~234, LP~717-36, GJ~3322), and in the secondary of
  the triple system GJ~852. We also confirm strong $2$~kG magnetic field
  in the primary of the triple system GJ~2005. 
  No fields could be accurately determined
  in rapidly rotating stars with $\vsini>10$~\kms. For slow and
  moderately rotating stars we find the surface magnetic field
  strength to increase with the rotational velocity $\vsini$ which is
  consistent with other results from studying field stars.  }
\end{abstract}


\section{Introduction}
\label{sec:intro}
Stellar rotation plays an important role in the evolution of single and
multiple systems, influencing the internal structures of stars, mixing
processes, surface temperature-pressure structure, convection, and, of
course, generation of the magnetic fields in subphotospheric layers where
strong differential rotation takes place.  In case of Sun-like and
low-mass stars, rotational velocity was found to decreases with time
\citep{1972ApJ...171..565S,2007ApJ...669.1167B}. Furthermore, this
evolution of rotation seems to work differently for dM stars with
fully convective envelopes.  These objects are apparently not braked
that much, and field dwarfs of spectral type late-M or L were found to
rotate more rapidly than their higher mass siblings
\citep{2003ApJ...583..451M, r-and-b-2008}.

The loss of angular momentum is caused by the stellar wind which is
driven by shock waves generated in atmospheres of solar type and cooler
stars where the surface convection is strong enough to
induce ultrasonic motions.  On the other hand, the efficiency of mass
loss probably depends on the intensity and geometry of the surface
magnetic field
\citep{mestel-1984,1988ApJ...333..236K,2000ApJ...534..335S}. These
fundamental characteristics of the magnetic field are determined by
the mechanism of its generation. Around spectral type M3.5, stars are
believed to become fully convective and thus the dynamo mechanism must
be different in cooler objects because they do not possess a
tachocline layer with strong differential rotation. Therefore, if the
magnetic field geometry changes in fully convective stars, a change in
the braking law may appear as well. An observed lack of slowly rotating
objects of spectral type later than mid-M could be a consequence of
such a change in the net braking. A problem in the interpretation of
stellar observations is that ``in general'' ages and other parameters like
the angle of inclination (important in observations of $\vsini$)
are unknown, at least in field stars.

Close double and hierarchical multiple system, however, are ideal
probes to study the rotational evolution of late-M objects. Their
components are most likely coeval and their spin axis aligned. Disc
orientations in pre-main sequence stars \citep{2006A&A...446..201M}
and orbit orientations in multiple systems \citep{2002A&A...384.1030S}
are partially correlated, and \citet{2000MNRAS.317..773B} find that
strong misalignments are unlikely in binaries with separations
$\leqslant 100$~AU. Thus, we may assume that the inclination angles
$i$ of close binary and multiple systems are near parallel. This
allows direct insight in spectral type dependent rotational braking
and its connection to magnetic field strength.

A first study of this kind was recently published by
\citet{2007A&A...471L...5R} based on the analysis of the close system
LHS~1070 (GJ~2005) composed of a mid-M star (M5.5) and two fainter
components \cb\ and \cc\ with spectral types around M9.0. Measuring
the astrometric orbit of \cb\ and \cc, \citet{2000A&A...353..691L}
determined masses at the limit to brown dwarfs. The full orbital
solution of the system shows both orbits of this triple system to be
co-planar within 2 degrees \citep{2008A&A...484..429S},
confirming the orbital alignment. It was found that magnetic flux in
the \cb\ component is about twice as strong as in component \cc\ at
similar rotation rate. \citet{2007A&A...471L...5R} concluded that rotational braking is
probably not proportional to magnetic field strength in fully
convective objects, and that a different field topology could be the
reason for the observed weak braking.

Motivated by the work of \citet{2007A&A...471L...5R}, we present an
analysis of seven carefully selected binary systems and two multiple
systems. Since the all components of multiple
systems are results of the same star formation process with presumably
identical age and metallicity, this allows us to study an evolution of
the rotational velocities from a well defined sample of stars.  Using
high-resolution VLT/CRIRES observations of FeH lines at
$\ll\,9920-9970$ we attempt to measure rotational velocities $\vsini$
and mean surface magnetic field $\bs$ in primary and secondary
components and to search for the links between rotational braking,
spectral type, and intensity of the surface magnetic field.

\section{Observations}
\label{sec:obs}

\begin{table*}
\renewcommand{\thefootnote}{\alph{footnote}}
\caption{M-dwarf systems.}
\label{tab:obs}
\begin{minipage}{\linewidth}
\begin{center}
\begin{tabular}{lcrcrcr}
\hline
\\
Name                      & Spectral type      & SNR & Separation, $\rho \pm 0.1$, arcsec      & Distance, pc & Age (Young?) & Reference\\
\\
\hline                    
\multirow{3}{*}{GJ~852}   & \ca\,\, M4.0     & 260 & \multirow{2}{*}{7.8 (\ca-\cc)} &&& \\
                          & \cb\,\, M4.5     & 170 & \multirow{2}{*}{0.9 (\cb-\cc)} & 10.0 &no   &   1, 3\\
                          & \cc\,\, M7.0\footnote{No spectral type was available in the
literature. Spectral types are determined based on the spectral type of
the primary and the near-infrared magnitude difference of the components.}     & 40 &                &&& \\
\hline
\multirow{2}{*}{GJ~4368 (LHS~4022)}                   & \ca\,\, M4.0     & 300 & \multirow{2}{*}{1.3}           & \multirow{2}{*}{11.0}&  \multirow{2}{*}{no} & \multirow{2}{*}{1, 4}\\
                & \cb\,\, M5.5     & 140 &                & &&\\
\hline
\multirow{2}{*}{LTT 7419} & \ca\,\, M2.5     & 250 &\multirow{2}{*}{15.5}           & \multirow{2}{*}{10.0\footnote{No distance
measurement available in the literature. Spectro-photometric distance
estimate is based on the V-K color of the primary which agrees with the
distance obtained from the R-K color of the secondary, assuming a spectral
type of M7.}} & \multirow{2}{*}{no} &     \multirow{2}{*}{5}\\
                          & \cb\,\, M7.0\footnotemark[1]     & 20  &                & &&\\
\hline
\multirow{2}{*}{GJ 234}   & \ca\,\, M4.5     & 100 & \multirow{2}{*}{1.0}           & \multirow{2}{*}{4.1} & \multirow{2}{*}{no} & \multirow{2}{*}{6, 7}\\
                          & \cb\,\, M7.0     & 100 &                & &&\\
\hline
\multirow{2}{*}{LP 717-36}& \ca\,\, M3.5     & 120 & \multirow{2}{*}{0.5}           & \multirow{2}{*}{20.2}&\multirow{2}{*}{yes} & \multirow{2}{*}{2, 3, 8, 9}\\
                          & \cb\,\, M4.0     & 100 &                & &&\\
\hline
\multirow{2}{*}{GJ 3322 (LP~476-207)}                  & \ca\,\, M4.0     & 170 & \multirow{2}{*}{1.2}           & \multirow{2}{*}{32.1}&\multirow{2}{*}{yes} & \multirow{2}{*}{10, 11, 12}\\
              & \cb\,\, M5.0\footnotemark[1]     & 90  &                & &&\\
\hline
\multirow{2}{*}{GJ 3304}  & \ca\,\, M4.0     & 70  & \multirow{2}{*}{0.8}           & \multirow{2}{*}{14.5}&\multirow{2}{*}{yes} & \multirow{2}{*}{1, 4, 8, 9}\\
                          & \cb\,\, M4.5     & 50  &                & &&\\
\hline
\multirow{2}{*}{GJ 3263 (LHS~1630)}                  & \ca\,\, M3.5     & 120 & \multirow{2}{*}{0.9}           & \multirow{2}{*}{13.3}&\multirow{2}{*}{no}  & \multirow{2}{*}{1, 4}\\
                & \cb\,\, M4.0\footnotemark[1]     & 80  &                &&& \\
\hline
\multirow{2}{*}{GJ 3344\footnote{Double-lined binary, components could not be spatially disantengled.} (LHS 1749)}                  & \ca\,\, M2.0              &  \multirow{2}{*}{110}   & \multirow{2}{*}{$\sim$}         & \multirow{2}{*}{$\sim$} &\multirow{2}{*}{$\sim$}&\multirow{2}{*}{$\sim$}\\
                & \cb\,\, M5.0              &     &                & &&\\
\hline
\multirow{3}{*}{GJ~2005 (LHS~1070)}   & \ca\,\, M5.5     & 50  & \multirow{2}{*}{1.35 (\ca-\cb)} & \multirow{3}{*}{7.7}& \multirow{3}{*}{no} &  \multirow{3}{*}{13, 14, 15}\\
 & \cb\,\, M8.5     & 20  & \multirow{2}{*}{1.75 (\ca-\cc)} & &&\\
                          & \cc\,\, M9.0     & 20  &                & &&\\
\hline
\end{tabular}
\end{center}
\end{minipage}
\begin{flushleft}
References for spectral types, distances, and youth:\\
1 - \citet{2004A&A...425..997B}, 2 - \citet{2004AJ....128..463R}, 3 - \citet{2006AJ....132..866R}, 4 - \citet{1997AJ....113.1458H},
5 - \citet{2003AJ....125..332J}, 6 - \citet{1991ApJS...77..417K}, 7 - \citet{1995msct.book.....B}, 
8 - \citet{2007ApJ...654..558D}, 9 - \citet{2009ApJ...699..649S}, 10 - \citet{1999A&A...344..897D}, 
11 - \citet{2003ApJ...599..342S}, 12 - \citet{2009AJ....137.4109L}, 13 - \citet{2000A&A...353..691L},
14 - \citet{2005AJ....130..337C}, 15 - \citet{2007A&A...471L...5R}.
\end{flushleft}
\end{table*}

The data were obtained in service mode between July and September 2008
with the NIR spectrograph CRIRES mounted at UT1 of the VLT (programme ID
81.D-0189). Observations followed a simple ABBA on-slit nod scheme,
collecting 4--10 spectra with individual exposure times of 180--300s,
depending on the brightness of the individual targets. For all
observations, both binary components were carefully aligned in the
entrance slit and nod throws were chosen to avoid overlaps of both sources
between A and B beams.

Data reduction followed the standard scheme of non-linearity correction of
all raw frames, pairwise subtraction of the science frames to remove the
sky background and flatfielding. 1D spectra were extracted from the
individual flatfielded and sky subtracted frames using an optimum
extraction algorithm and by carefully masking the nearby component to
exclude flux contamination in those close double stars. Due to the
inherent slit curvature of CRIRES, spectra taken in different nodding
positions do not share the same wavelength scale. To correct for this
effect we calculated individual wavelength solutions. We then mapped all
1D spectra onto a common wavelength grid and finally co-added all rebinned
spectra of each component. The wavelength solutions are based at first
order on the ThAr calibration frames provided by ESO and were refined by
matching the science spectra to the atlas of GJ1002, previously observed
with CRIRES \citep{wende-2010}.

Table~\ref{tab:obs} summarises some information about the observed systems, including
the observed separation of the components in our CRIRES spectra as well as
the distance to the system which is mostly based on trigonometric
parallaxes. In addition, we indicate which systems are identified in the
literature as being younger than 300~Myrs, based on their membership in a
young moving group or by showing low surface gravity and strong X-ray and
$\halpha$ activity.

We also extended the original sample of stars by GJ~2005, a triple system
analysed previously by \citet{2007A&A...471L...5R}. This system was also
observed with CRIRES (programme ID 60.A-9078). Unfortunately, the
components of the system GJ~3344 (LHS~1749) could not be spatially
disentangled due to strong orbital motion between the discovery of this
binary and our observations. We could thus not perform individual analysis
of each of the components and excluded the system from our investigation.
For all other systems, we obtained individual spectra of components,
though the signal-to-noise ratio is very low for some of them (mostly for
the fainter \cb\ and \cc\ components).

\section{Methods}
\label{sec:methods}

\subsection{Input line lists and synthetic spectra}
In our investigation we employed the FeH line list of the Wing-Ford band ($F^4\,\Delta-X^4\,\Delta$ transitions
at $1\,\mu$m) and molecular constants taken from \citet{dulick2003}\footnote{http://bernath.uwaterloo.ca/FeH/}.
For some of these lines we used corrected Einstein~A values according to \citet{wende-2010}.

To compute synthetic spectra of the atomic and molecular lines in
the presence of a magnetic field, we employed the \synmast\ code \citep{synthmag2007}. The code
represents an improved version of the \synthmag\ code described by
\citet{synthmag}. It solves the polarized radiative transfer equation for a
given model atmosphere, atomic and molecular line lists and magnetic field
parameters. The code treats simultaneously thousands of blended absorption
lines, taking into account their individual magnetic splitting patterns, which
can be computed for the Zeeman or the Paschen-Back regime. \synmast\ provides 
local four Stokes parameter spectra for a number of angles between the
surface normal and the line of sight ($7$ by default). These local spectra are
convolved with appropriate rotational, macroturbulent and instrumental profiles
and then combined to simulate the stellar flux profiles.

Model atmospheres are from the recent MARCS grid\footnote{http://marcs.astro.uu.se} \citep{marcs}.

\subsection{Molecular Zeeman effect}
In order to analyse the magnetic field through spectral synthesis it is necessary to know
the Land\'e g-factors of upper and lower levels of a particular molecular transition.
As to diatomic molecules, a simple analytical expression for g-factors
can be obtained only in special cases of pure Hund's \textit{A} and \textit{B} splitting of energy levels.
These cases correspond to the strong and weak coupling of the 
electronic spin $\mathbf{S}$ and orbital $\mathbf{L}$ momenta to the internuclear axis
(i.e. coupling with nuclear rotation) \citep[see][]{herzberg1950}. 
Unfortunately, as stated in \citet{berdyugina2002}, 
the lines of FeH of the Wing-Ford band exhibit splitting 
that is in most cases intermediate between pure Hund's \textit{A} and \textit{B} 
and which is not trivial to treat both theoretically and numerically.
Available theoretical descriptions make use of the approach based on the
so-called effective Hamiltonian that is usually represented as a sum of the unperturbed part
describing energies of Zeeman levels as they undergo transition between Hund's cases, 
and the part describing an interactions with the external
magnetic field. A detailed description can be found, for example, in \citet{berdyugina2002} and \citet{ramos2006}.

To compute g-factors we implement numerical libraries from the MZL (Molecular Zeeman Library)
package originally written by B.~Leroy \citep{mzl}, and adopted
for the particular case of FeH. MZL is a collection of routines for computing the Zeeman effect in diatomic molecules,
and it contains all the physics of pure and intermediate Hund's cases presented in \citet{berdyugina2002}.
Due to the limitations of the theoretical description of the intermediate Hund's case
presented in \citet{berdyugina2002}, the calculated g-factors in most cases fail to fit the observed
splitting of FeH lines. This has already been mentioned in a number of previous investigations
\citep[see][]{afram2008,berdyugina2002,harrison-brown2008}. Therefore, in the present investigation
we used an alternative approach described in \citet{2010A&A...523A..37S}. The authors suggested a semi-empirical
method of computing g-factors using MZL routines and different Hund's cases for upper and lower energy levels
of particular FeH transitions depending on their quantum numbers and the fit to the spectra of a sunspot
with known magnetic field intensity. This approach has been successfully applied to the analysis of the
magnetic field in FeH spectra of several magnetic M-dwarfs and we refer the interested reader
to the original work by \citet{2010A&A...523A..37S}.

\subsection{Analysis of the spectra}
\label{sec:analysis}
The spectroscopic estimation of the atmospheric parameters for each system is a complicated task because of
the short wavelength region $\ll\,9920-9972$ observed with CRIRES.
In spite of the fact that, in general, lines of FeH are good indicators of the atmospheric temperature 
and the magnetic field strength,
no lines of other species (especially atoms) are present in this spectral range that could be used for cross-checking.
Moreover, the van der Waals damping constants $\gammaw$ are unknown for FeH transitions.
In case of slow rotation, pressure broadening is the dominating broadening mechanism in the dense plasma of dM stars,
and thus the knowledge of $\gammaw$  is of high importance when it turns to quantitative analysis of individual line profiles.
As found by \citet{2010A&A...523A..37S} and \citet{wende-2010}, the classical $\gammaw$ \citep{gray} must
be increased by a factor of $\approx3.5$ to fit FeH lines in the spectra of non-magnetic M5.5 dwarf GJ~1002.
Note that any uncertainties in $\gammaw$ would immediately affect
the estimation of other parameters, such as $\vsini$, Fe abundance $\abn{Fe}$, and $\teff$. Hence, when analysing
spectra of M-dwarfs, some of these key parameters must be known independently and prior to spectroscopic fitting.
However, there is a strong evidence that $\gammaw$ drops significantly for late M's, and its value
must be estimated using spectra of some non-magnetic stars. Here we made use of spectra of the weakly magnetic M8.0 star VB~10,
but its rather high $\vsini=6$~\kms\ \citep{r-and-b-2007} makes it still difficult to precisely determine the FeH damping constant.
Also, the analysis of Keck/HIRES spectra of three M8V dwarfs from a 
program of \citet{2010ApJ...710..924R} did not warrant accurate results, mainly because
of low resolution of the data (see Sect.~\ref{sec:results:vdw} for details).
We therefore investigated for every system a possible range of atmospheric parameters under different assumptions
about $\gammaw$: since any uncertainties in $\teff$ or $\gammaw$ immediately transform into uncertainties of
$\vsini$ and $\abn{Fe}$ (and vice verse), it is essential to fix 
either $\teff$ or $\abn{Fe}$ and then search for the combination of remaining parameters
($\gammaw$ and $\vsini$) that provides the best fit. Usually, spectral types are known from photometric
calibrations. Here we utilise the effective temperature scale given in \citet{1995ApJS..101..117K} for stars
in the spectral type range [M0.0:M6.0] and from \citet{2004AJ....127.3516G} for [M6.5:M9.5] respectively. 
The reference abundance of iron is assumed to be solar $\abn{Fe}=-4.59$ \citep{asplund2005}.
We also assume $\logg=5.0$ for all objects in our sample which is the usual value
for dM stars \citep{2003A&A...397L...5S,2006ApJ...652.1604B,2006ApJ...644..475B}.

Finally, the basic steps of spectroscopic analysis can be summarized as follows:
\begin{enumerate}
\item
Using a set of magnetically insensitive FeH lines we derive fitting parameters
such as $\abn{Fe}$, $\vsini$, $\fgammaw$ (enhancement factor by which the classical $\gammaw$ 
is multiplied), and $\teff$ by least-square minimisation fit. 
The magnetic insensitive lines used are: 
FeH $\ll\,9941.62$, $9944.56$, $9945.83$, $9953.07$, $9957.30$, $9962.83$.
\item
The intensity of the surface magnetic field $\bs$ was determined by constructing
a $\chi^2$ goodness-of-the-fit landscape of the deviation between observed and predicted FeH spectra for a number of
values of the magnetic field intensity.
\item
In case of very fast rotating stars all the available spectral range of $\ll\,9920-9972$ was fitted 
because of strong line blending. No accurate estimate of the magnetic field was possible for such objects.
\end{enumerate}

To provide a comprehensive look into the possible parameter space of each stellar component, 
the following compilations of fitted parameters were considered:
\begin{enumerate}
\item
determining $\vsini$ with fixed $\teff$, $\abn{Fe}$, and $\fgammaw$;
\item
determining $\teff$ and $\vsini$ with fixed $\abn{Fe}$ and $\fgammaw$;
\item
determining $\fgammaw$ and $\vsini$ with fixed $\teff$ and $\abn{Fe}$.
\end{enumerate}

Using this approach we find rotational velocities
and surface magnetic field intensities (where possible) using the \synmast\ code and Land\'e g-factors calculated
as described in \citet{2010A&A...523A..37S}. In the next section we present detailed results of our investigation.

\section{Results}
\label{sec:results}

\subsection{Pressure broadening of FeH lines}
\label{sec:results:vdw}

Based on investigations of FeH lines in spectra of the non-magnetic and
slowly rotating M5.5 dwarf GJ~1002 it was shown that to get a
satisfactory agreement between observed and predicted spectra one
needs a $3.5$ times increase of the classical van der Waals damping
constant $\gammaw$ \citep[see][]{2010A&A...523A..37S,wende-2010}.  To
verify the behaviour of $\gammaw$ in much cooler plasma we analysed
spectra of late type non- or weakly active M-dwarfs 
2MASS-1246517+314811 (M7.5, $\teff=2550$~K),
2MASS-1440229+133923 (M8.0, $\teff=2500$~K),
2MASS-2037071-113756 (M8.0, $\teff=2500$~K) from
\citet{2010ApJ...710..924R}, and VB~10 (M8.0, $\teff=2500$~K).  
Assuming effective temperatures to be known
from stellar spectral types and solar iron abundance, the fit to the
observed spectra in all late M's resulted in a strong decrease of
$\gammaw$.  For instance, for 2MASS-2037 an optimal fit is
obtained with $\fgammaw\approx0.16$ and $\vsini=4$~\kms. Decreasing
$\vsini$ to $1$~\kms\ results in only in marginal increase of
$\fgammaw$.  Even smaller $\fgammaw\approx0.06$ is found for
2MASS-1440 and 2MASS-1246. Thus, there is an
unexpected difference in $\gammaw$ for these three stars as it should
be (naively) lower for 2MASS-2037 than for the hotter
2MASS-1246, or at least nearly the same. We speculate that
the explanation can be connected with a) a slightly different iron
abundance in the atmospheres of these objects, b) uncertainties in
spectral type determination, and/or c) the quality of the spectra (SNR).

\begin{figure*}
\centerline{\includegraphics[width=0.7\hsize]{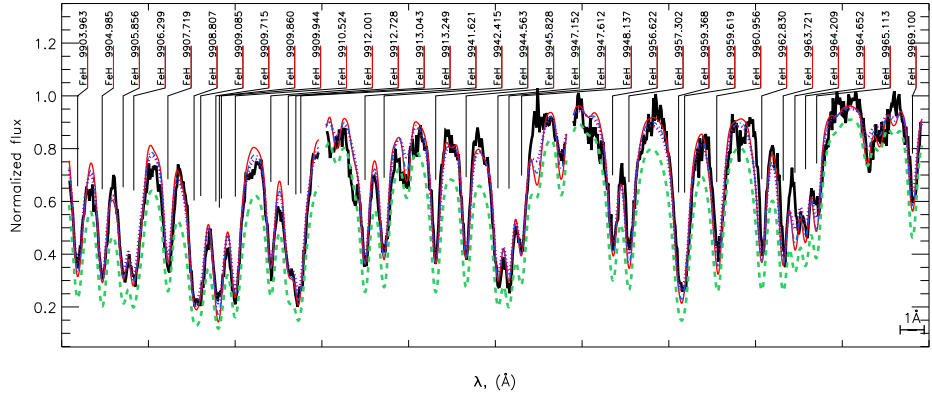}}
\caption{Comparison between observed and predicted profiles of FeH lines in spectra of M8.0 dwarf VB~10.
Thick line~--~observations; thin full line (red)~--~$\teff=2500$~K, $\fgammaw=0.3$;
dashed line (green)~--~$\teff=2500$~K, $\fgammaw=1.0$;
dash-dotted line (blue)~--~$\teff=2600$~K, $\fgammaw=0.8$,
dotted line (violet)~--~$\teff=2700$~K, $\fgammaw=2.0$. For all models
solar Fe abundance and $\vsini=6$~\kms\ were used.}
\label{fig:vb10}
\end{figure*}

The same conclusion follows from the analysis of the spectra of VB~10,
which has the highest resolution of $R\approx85\,000$ among the four
targets, but, on the other hand, higher rotational velocity.
Assuming fixed $\vsini=6$~\kms\ from \citet{r-and-b-2007} results in
$\fgammaw=0.3\pm0.01$, i.e. even larger than the value found for
2MASS-2037. Fixing $\fgammaw=3.5$ needs a substantial
decrease of $\vsini=1.18$~\kms\ and $\abn{Fe}=-5.28$, however with
slightly better fit.  Increasing $\teff$ from $2500$~K to $2600~K$ and
fixing solar $\abn{Fe}=-4.59$ and $\vsini=6$~\kms\ gives again better
fit but larger $\fgammaw=0.74$ and $\gammaw\approx2$ for even higher
$\teff=2700$~K, respectively.  Note that the error bars for spectral
type of M-dwarfs are usually $\pm0.5$ which translates to a minimum error in
temperature of approximately $\pm50$~K. Therefore, $\teff=2700$~K for VB~10
seems not well justified, and thus $\gammaw$ is likely to have values $<1.0$ for
late-type objects, but its true value is difficult to determine at the present stage. 
As an example, Fig.~\ref{fig:vb10} illustrates theoretical fits to some of FeH lines.

As mentioned above, we assumed the same $\logg=5.0$ for all early and late type dM stars.
The later ones, however, may have larger $\logg$ values 
\citep[see, e.g.,][]{2003A&A...397L...5S,2006ApJ...652.1604B,2006ApJ...644..475B} that can reach
up to $5.4$~dex at spectral type M8.0. Obviously, this cannot help to resolve the issue with decrease
in $\fgammaw$ for the coolest M dwarfs simply because higher $\logg$ would broaden profiles of spectral lines,
thus requiring even smaller $\fgammaw$ then those found here.

Taking into account a certain difficulty to accurately estimate the pressure broadening constant, we thus
examine for each binary system a set of solutions depending on assumed value of $\fgammaw=[0.3,1.0,3.5]$. 
This allowed us to estimate a possible impact of the uncertainty in $\teff$, $\fgammaw$, and $\abn{Fe}$ on final results 
(e.g., $\vsini$ and $\bs$). Note that in case of early type M stars we use mainly $\fgammaw=[1.0,3.5]$.

\begin{table*}
\caption{Atmospheric parameters of M-dwarf systems, obtained under the assumption of fixed $\abn{Fe}$ and $\fgammaw$ (Case~2).}
\begin{footnotesize}
\begin{center}
\begin{tabular}{+l^c^c^r^r^r^r^r^r^c}
\hline
System                               & Component          & Spectral            & $\teff^{\rm (0)}$  & $\teff$     & $\abn{Fe}$   & $\vsini$    & $\fgammaw$  & $\b$ & $\chi^2$\\
                                     &                    &  type               & (K)              & (K)         &              & (\kms)      &             & (kG) &  (GOF)  \\
\hline
\multirow{3}{*}{GJ~852}              & \multirow{1}{*}{A} & \multirow{1}{*}{M4.0} &3366&  3133    & -4.59        & 5.14     & 3.5         & 3      & 0.2836  \\
                                     & \multirow{1}{*}{B} & \multirow{1}{*}{M4.5} &3310&  3132    & -4.59        & 3.66     & 3.5         & 1.6    & 0.1644  \\
                                     & \multirow{1}{*}{C} & \multirow{1}{*}{M7.0} &2621&  2582    & -4.59        & 11.38    & 0.3         & $\sim$ & 2.1322  \\
\hline                                                                                                                                     
\multirow{2}{*}{GJ~4368 (LHS~4022)}  & \multirow{1}{*}{A} & \multirow{1}{*}{M4.0} &3366&  3554 & -4.59        & 3.01  & 3.5        & 0.5  & 0.0662  \\
                                     & \multirow{1}{*}{B} & \multirow{1}{*}{M5.5} &3250& 3375   & -4.59        & 2.80  & 3.5         & 0.79  & 0.1206  \\
\hline                                                                                                                                     
\multirow{2}{*}{LTT~7419}            & \multirow{1}{*}{A} & \multirow{1}{*}{M2.0} &3525&  3741   & -4.59        & 3.62  & 3.5         & 0    & 0.0551  \\
                                     & \multirow{1}{*}{B} & \multirow{1}{*}{M8.0} &2621&  2681  & -4.59        & 12.35 & 0.3         & $\sim$  & 2.8455  \\
\hline                                                                               
\multirow{2}{*}{GJ~234}              & \multirow{1}{*}{A} & \multirow{1}{*}{M4.5} &3310&  3152  & -4.59        & 6.15  & 3.5         & 2.75 & 0.1731  \\
                                     & \multirow{1}{*}{B} & \multirow{1}{*}{M7.0} &2621&  2737  & -4.59        & 12.93 & 0.3         & $\sim$ & 0.8577  \\
\hline                                                                               
\multirow{2}{*}{LP~717-36}          & \multirow{1}{*}{A} & \multirow{1}{*}{M3.5} &3418&  3346  & -4.59        & 3.94  & 3.5         & 1.75 & 0.1123  \\
                                     & \multirow{1}{*}{B} & \multirow{1}{*}{M4.0} &3366& 3163  & -4.59        & 12.16 &   3.5       &  $\sim$   & 0.6185  \\
\hline                                                                               
\multirow{2}{*}{GJ~3322 (LP~476-207)}& \multirow{1}{*}{A} & \multirow{1}{*}{M4.0} &3366&  3289  & -4.59        & 6.75   &  3.5        & 3    & 0.1200  \\
                                     & \multirow{1}{*}{B} & \multirow{1}{*}{M5.0} &3250&  3101  & -4.59        & 23.77 &   3.5       &  $\sim$   & 0.4277  \\
\hline                                                                               
\multirow{2}{*}{GJ~3304}             & \multirow{1}{*}{A} & \multirow{1}{*}{M4.0} &3366&  3083  & -4.59        & 29.73 &   3.5       &  $\sim$   & 0.3142  \\
                                     & \multirow{1}{*}{B} & \multirow{1}{*}{M4.5} &3310&  3084  & -4.59        & 45.35 &   3.5       &  $\sim$ & 0.2956  \\
\hline                                                                               
\multirow{2}{*}{GJ~3263 (LHS~1630)}  & \multirow{1}{*}{A} & \multirow{1}{*}{M3.5} &3418&  3600   & -4.59        & 2.27  &  3.5        &  0.75& 0.0930  \\
                                     & \multirow{1}{*}{B} & \multirow{1}{*}{M4.0} &3366&  3518   & -4.59        & 2.73  &  3.5        &  0.5 & 0.1178  \\
\hline                                                                                                                                     
\multirow{3}{*}{GJ~2005 (LHS~1070)}  & \multirow{1}{*}{A} & \multirow{1}{*}{M5.5} &3160&  2982    & -4.59        &  6.82    & 3.5         & 2    & 0.1541 \\
                                     & \multirow{1}{*}{B} & \multirow{1}{*}{M8.5} &2450&  $<$2500 & -4.59        & 15.83    & 0.3         & $\sim$ &  5.5399  \\
                                     & \multirow{1}{*}{C} & \multirow{1}{*}{M9.0} &2400&  $<$2500 & -4.59        & 15.06    & 0.3         & $\sim$ &  5.9756  \\
\hline
\end{tabular}
\end{center}
\begin{flushleft}
$\teff^{\rm (0)}$ is the temperature which corresponds to the spectral type of the star. \\
No accurate estimates of magnetic field were possible for stars with $\vsini>10$~\kms\ (marked with ``$\sim$'') \\
Last column refer to the chi-square goodness of the fit (GOF) values. \\
For GJ~2005 only upper limit of $\teff$ can be derived
because of temperature limitation of the available MARCS model grid.
\end{flushleft}
\end{footnotesize}
\label{tab:param-0-0-1-1}
\end{table*}

\subsection{Atmospheric parameters of stars in dM systems}

In this paragraph we present estimates of the $\vsini$ and $\bs$ obtained 
for three basic sets of fitted parameters introduced in Sect.~\ref{sec:analysis}.

\subsubsection*{Case~1: determining $\vsini$ and $\bs$ with fixed $\teff$, $\abn{Fe}$, and $\fgammaw$}

Formally, the easiest and the straightforward way of the analysis would be to assume that $\teff$ is known from
the spectral type assigned to each star. Then, the iron abundance would be solar or nearly solar as all objects belong
to solar neighborhood and their atmospheres are well mixed due to strong convection. The only issue left is the $\fgammaw$,
which value is most uncertain. Nevertheless, it can be taken from recent investigations \citep{2010A&A...523A..37S,wende-2010}
for early-type objects and from the analysis given in Sect.~\ref{sec:results:vdw} of this work for late-type ones.
Theoretical computations showed, however, that fixing $\teff$, $\abn{Fe}$, and $\fgammaw$ always resulted in the worst fit
for all objects. Furthermore, setting $\fgammaw=1.0$ also did not improve the fit for both early- and late-type stars. 
This indicated that some of the fixed parameters must differ from assumed values, and in the next step we tried to
explore how much could these differences be and what is their impact on estimated $\vsini$ and $\bs$.

\subsubsection*{Case~2: determining $\teff$, $\vsini$, and $\bs$ with fixed $\abn{Fe}$ and $\fgammaw$}

\begin{figure*}
\centerline{
\includegraphics[width=0.5\hsize]{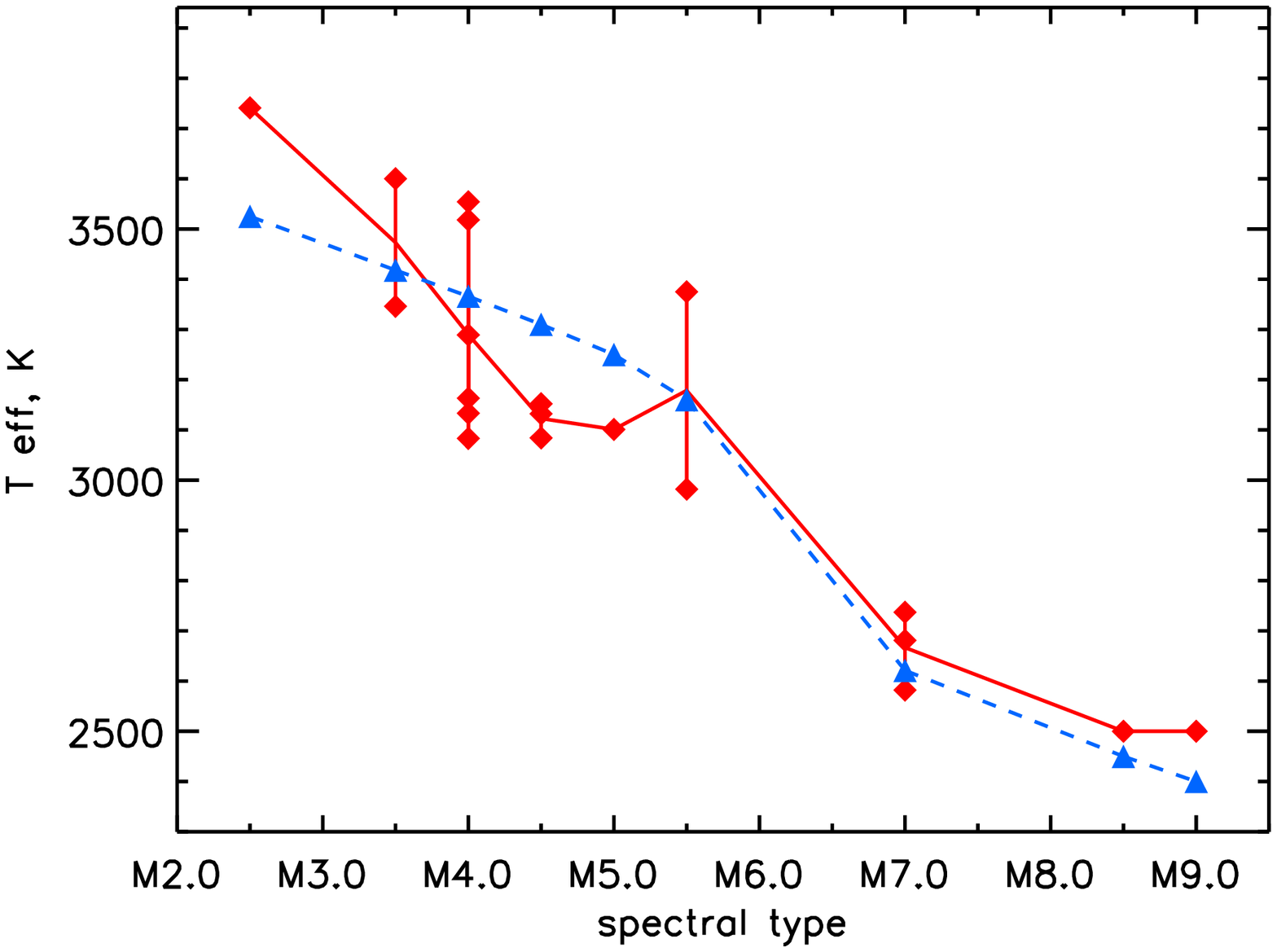}
\includegraphics[width=0.5\hsize]{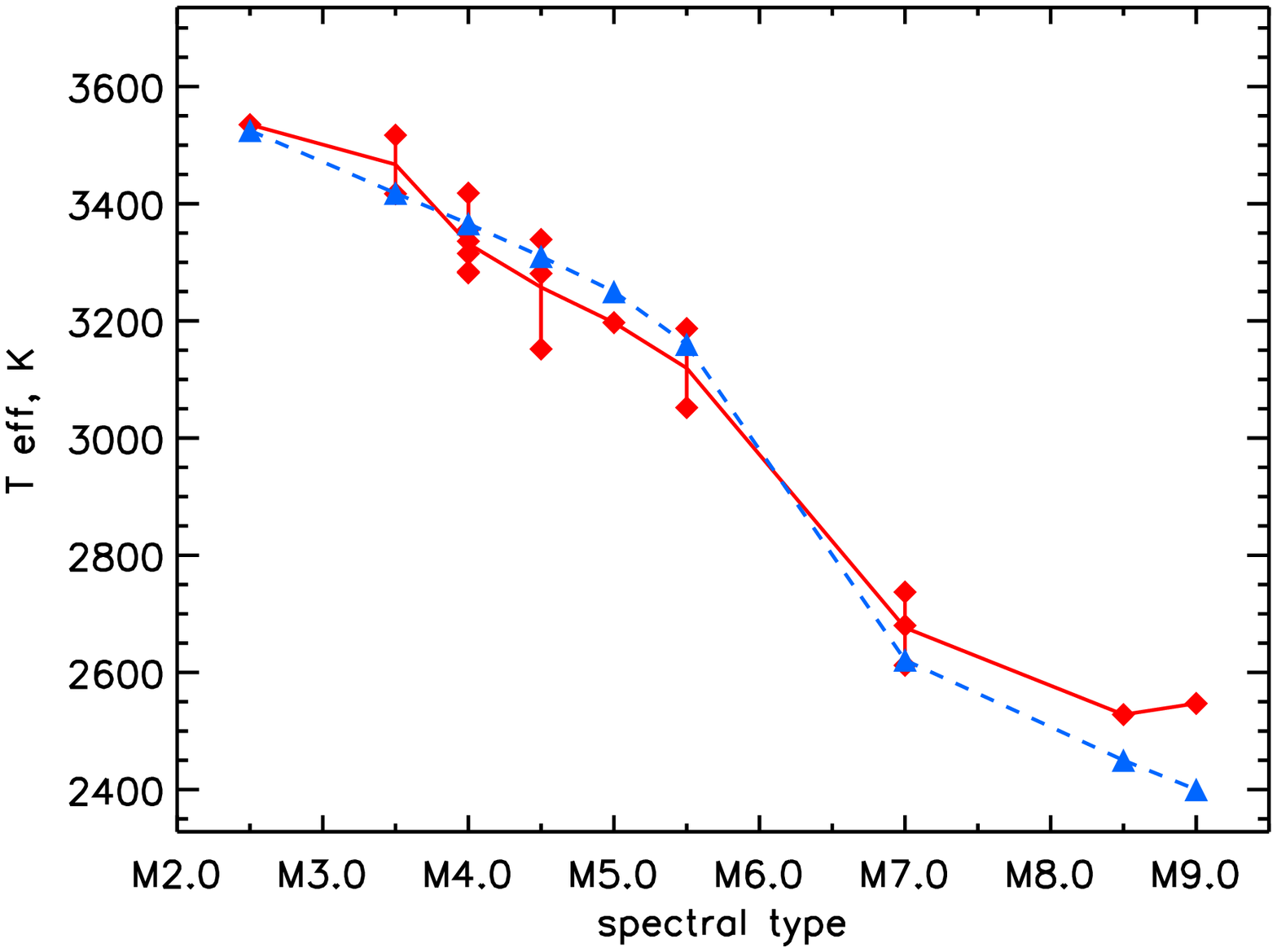}
}
\caption{
Run of the effective temperature as a function of spectral type derived from the FeH lines for
Case~2 (\textbf{left panel}) and Case~4 (\textbf{right panel}). See text for more details.
Red diamonds represent measurements for each individual object and solid line connects their mean values.
Blue dashed line correspond to empirical calibrations after \citet{1995ApJS..101..117K} and \citet{2004AJ....127.3516G}.
}
\label{fig:t-sp}
\end{figure*}

Keeping $\abn{Fe}$ fixed  at its solar value and $\fgammaw=[0.3,3.5]$ for late- and early-type stars respectively
allowed us to obtain a much better agreement
between observed and predicted profiles of FeH lines, and in many (but not all) cases this resulted in the best fit.
Table~\ref{tab:param-0-0-1-1} summarizes results for each system, and its inspection immediately leads to the
following conclusion: in all cases the obtained $\teff$ deviates 
from the temperature estimated from the individual spectral type using
calibrations after \citet{1995ApJS..101..117K} and \citet{2004AJ....127.3516G}.
Large deviations are found for GJ~3304-\ca~(283~K) and GJ~852-\ca~(233~K).
The deviation from the empirical $\teff$-spectral type relation and the one derived here is plotted on
the left panel of Fig.~\ref{fig:t-sp}. Apart from the big scatter in the derived $\teff$ even for objects of the same spectral type,
the run of $\teff$ roughly follows empirical predictions. 
We thus speculate that probably the spectral type was incorrectly assigned 
to some of our objects. Or, as will be shown below, the iron abundance
of stars may differ from the solar one which we assumed here.

\subsubsection*{Case~3: determining $\abn{Fe}$, $\vsini$, and $\bs$ with fixed $\teff$ and $\fgammaw$}

\begin{figure}
\centerline{\includegraphics[width=\hsize]{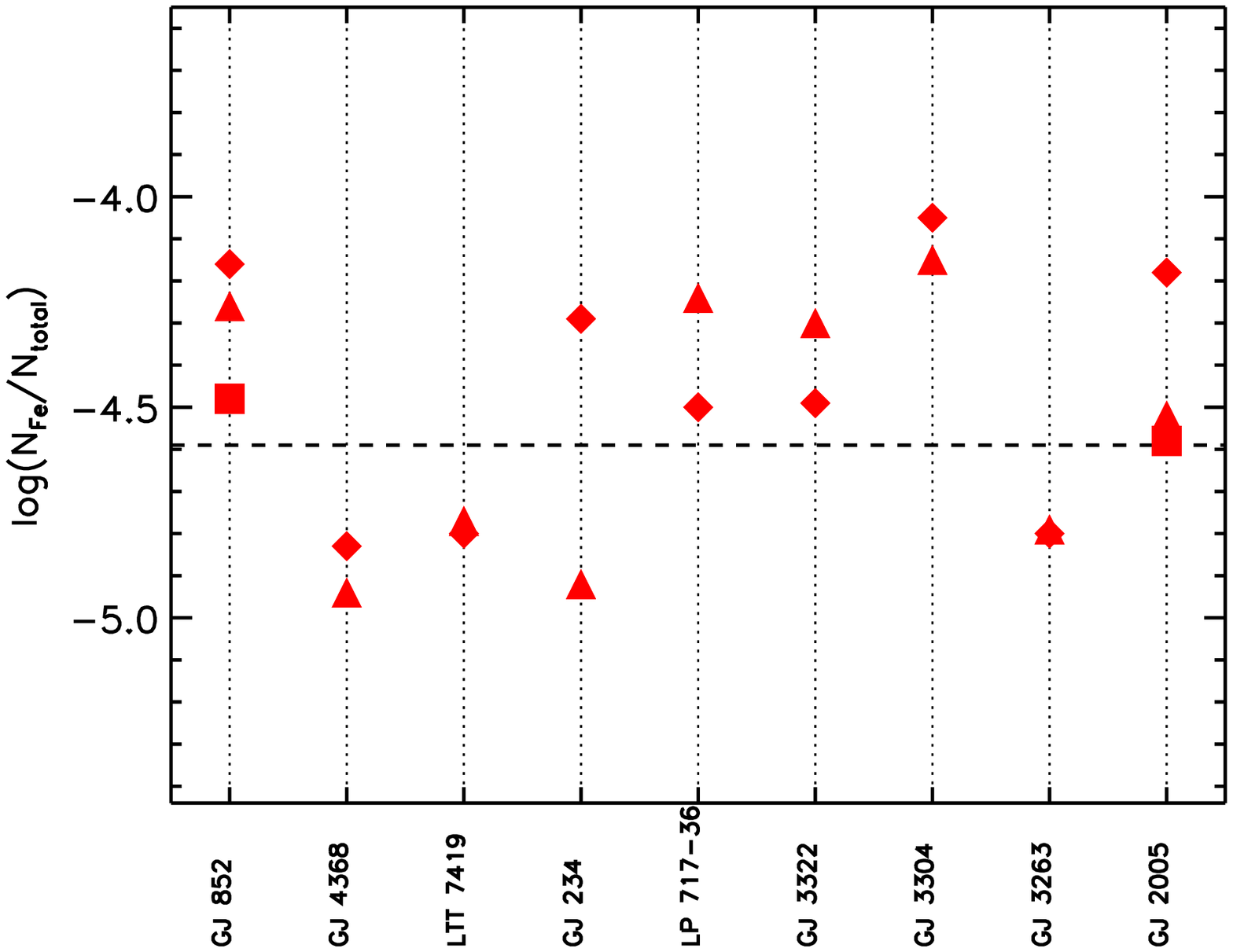}}
\caption{
Derived iron abundance for \ca~(diamonds), \cb~(triangles), and \cc~(squares) components of investigated dM systems. 
The reference solar abundance is shown by dashed horizontal line.
}
\label{fig:1-0-0-1-abn}
\end{figure}

This combination of fitted parameters also provided a very good
agreement between observed and predicted profiles of FeH lines and the
results are presented in Table~\ref{tab:param-1-0-0-1}.  In
particular, keeping $\teff$ fixed immediately reflected in the strong
changes of the iron abundance, which in some cases was found to be up
to $0.5$~dex lower or larger than its current solar value (e.g.
GJ~4368-B, GJ~3304).  Figure~\ref{fig:1-0-0-1-abn} shows derived iron abundance
for each star.  It is seen that for this case, $\abn{Fe}$ differs
significantly among objects of the same system, which is difficult to
understand from the physical point of view as naturally one would
expect that all the components have the same abundance patterns
due to their common origin and history. It is interesting that in most cases
$\abn{Fe}$ of the primary is higher than that of the secondary. This may be
a hint for an effect of the degeneracy between temperature and Fe
abundance in our models. Consistent results are found for LP~7419 and
GJ~3263. Values close to solar are found
for GJ~852-\cb, LP~717-36-\ca, GJ~3322-\ca, and GJ~2005-\cb\cc.

The most interesting outcome of fixing $\teff$ and $\fgammaw$, however, is that the derived $\vsini$ and $\bs$ does not change much,
and their values in Table~\ref{tab:param-0-0-1-1} and Table~\ref{tab:param-1-0-0-1} are in good agreement between each other.
This is an important result since it tells us that even if our estimates of abundances and/or temperatures are wrong by
any means, this has little influence on the derived rotational velocity and surface magnetic field. In case of rotation,
fixing $\teff$ at a certain value simply requires the adjustment of $\abn{Fe}$ in order to obtain minimum deviation between
observed and theoretical spectra. Note that fitting is done using magnetically insensitive FeH lines for slow and
moderately rotating stars and thus this in no way influences estimates of the magnetic field.
The estimated strength of the later is driven by Zeeman-broadened lines and thus should weakly depend on
assumed $\teff$ and $\abn{Fe}$, of course if rotation is relatively slow. This is exactly what is seen from 
Table~\ref{tab:param-1-0-0-1}.

\subsubsection*{Case~4: determining $\teff$, $\vsini$, and $\bs$ with fixed $\fgammaw$ and mean $\abn{Fe}$}

As a last step in our spectroscopic analysis we determined $\teff$ and
$\vsini$, but this time assuming iron abundance to be identical
for members of the same system and its values were simply taken as a
mean of individual values from the Case~3 (see Table~\ref{tab:param-1-0-0-1}). 
Table~\ref{tab:param-Fe_mean-0-0-1-1} summarizes the obtained
parameters. Again, and similar to previous case, neither rotational
velocities nor the surface magnetic field changed much. Using a mean
iron abundance, however, led to somewhat better agreement between
derived and empirically calibrated effective temperatures, as can be
seen from the right panel of Fig.~\ref{fig:t-sp}.  In particular,
temperatures of objects of types M3.5 and M5.0 get closer to empirical
values. Yet there still exist noticeable deviations between
predicted and empirical values.

\begin{table*}
\caption{Atmospheric parameters M-dwarf systems, obtained under the assumption of fixed $\teff$ and $\fgammaw$ (Case~3).}
\label{tab:param-1-0-0-1}
\begin{footnotesize}
\begin{center}
\begin{tabular}{+l^c^c^r^r^r^r^r^c}
\hline
System                               & Component          & Spectral              & $\teff$     & $\abn{Fe}$   & $\vsini$    & $\fgammaw$  & $\b$ & $\chi^2$\\
                                     &                    &  type                 & (K)         &              & (\kms)      &             & (kG) &  (GOF)  \\
\hline
\multirow{3}{*}{GJ~852}              & \multirow{1}{*}{A} & \multirow{1}{*}{M4.0} &  3366       & -4.16 &   5.43     & 3.5         & 3     & 0.4461  \\
                                     & \multirow{1}{*}{B} & \multirow{1}{*}{M4.5} &  3310       & -4.26 &   3.90     & 3.5        & 1.5    & 0.1749  \\
                                     & \multirow{1}{*}{C} & \multirow{1}{*}{M7.0} &  2621       & -4.48 &  11.35     & 0.3        & $\sim$ & 2.2511   \\
\hline                                                                                                                                     
\multirow{2}{*}{GJ~4368 (LHS~4022)}  & \multirow{1}{*}{A} & \multirow{1}{*}{M4.0} &  3366       & -4.83  & 2.74  & 3.5         & 0.5   & 0.0660  \\
                                     & \multirow{1}{*}{B} & \multirow{1}{*}{M5.5} &  3160       & -4.94  & 2.41  & 3.5         & 0.75  & 0.1249  \\
\hline                                                                                                                                     
\multirow{2}{*}{LTT~7419}            & \multirow{1}{*}{A} & \multirow{1}{*}{M2.5} &  3525       & -4.80  &  3.40 & 3.5         & 0       & 0.0523  \\
                                     & \multirow{1}{*}{B} & \multirow{1}{*}{M7.0} &  2621       & -4.77  & 12.49 & 0.3         & $\sim$  & 2.8463  \\
\hline                                                                               
\multirow{2}{*}{GJ~234}              & \multirow{1}{*}{A} & \multirow{1}{*}{M4.5} &  3310       & -4.29  &  6.35 & 3.5         & 2.5    & 0.2249  \\
                                     & \multirow{1}{*}{B} & \multirow{1}{*}{M7.0} &  2621       & -4.92  & 13.12 & 0.3         & $\sim$ & 0.7689  \\
\hline                                                                               
\multirow{2}{*}{LP~717-36}           & \multirow{1}{*}{A} & \multirow{1}{*}{M3.5} &  3418       & -4.50  &  3.98 & 3.5         & 1.75    & 0.1145  \\
                                     & \multirow{1}{*}{B} & \multirow{1}{*}{M4.0} &  3366       & -4.24  & 12.17 &   3.5       &  $\sim$ & 0.6142  \\
\hline                                                                               
\multirow{2}{*}{GJ~3322 (LP~476-207)}& \multirow{1}{*}{A} & \multirow{1}{*}{M4.0} &  3366       & -4.49  &  6.82 &   3.5       &  3   & 0.1356  \\
                                     & \multirow{1}{*}{B} & \multirow{1}{*}{M5.0} &  3250       & -4.30  & 23.80 &   3.5       &  $\sim$   & 0.3949  \\
\hline                                                                               
\multirow{2}{*}{GJ~3304}             & \multirow{1}{*}{A} & \multirow{1}{*}{M4.0} &  3366       & -4.05  & 29.19 &   3.5       &  $\sim$ & 0.2908  \\
                                     & \multirow{1}{*}{B} & \multirow{1}{*}{M4.5} &  3310       & -4.15  & 44.78 &   3.5       &  $\sim$ & 0.2642  \\
\hline                                                                               
\multirow{2}{*}{GJ~3263 (LHS~1630)}  & \multirow{1}{*}{A} & \multirow{1}{*}{M3.5} &  3418       & -4.80  & 2.10  &  3.5        &  0.75& 0.0922  \\
                                     & \multirow{1}{*}{B} & \multirow{1}{*}{M4.0} &  3366       & -4.79  & 2.52  &  3.5        &  0.5 & 0.1161  \\
\hline
\multirow{3}{*}{GJ~2005 (LHS~1070)}  & \multirow{1}{*}{A} & \multirow{1}{*}{M5.5} &  3160       & -4.18  &   6.89     & 3.5        & 1.75 & 0.1928 \\
                                     & \multirow{1}{*}{B} & \multirow{1}{*}{M8.5} &  2500$^*$   & -4.52  &  15.94     & 0.3        & $\sim$   & 5.7981  \\
                                     & \multirow{1}{*}{C} & \multirow{1}{*}{M9.0} &  2500$^*$   & -4.58  &  15.11     & 0.3        & $\sim$   &  5.9962   \\
\hline
\end{tabular}
\end{center}
\begin{flushleft}
No accurate estimates of magnetic field were possible for stars with $\vsini>10$~\kms\ (marked with ``$\sim$'') \\
Last column refer to the chi-square goodness of the fit (GOF) values.\\
($^*$)~--~coolest available model with $\teff=2500$ was used.
\end{flushleft}
\end{footnotesize}
\end{table*}

\begin{table*}
\caption{Atmospheric parameters M-dwarf systems, obtained under the assumption of fixed $\abn{Fe}$ and $\fgammaw$ (Case~4).}
\begin{footnotesize}
\begin{center}
\begin{tabular}{+l^c^c^r^r^r^r^r^r^c}
\hline
System                               & Component          & Spectral            & $\teff^{\rm (0)}$  & $\teff$     & $\abn{Fe}$   & $\vsini$    & $\fgammaw$  & $\b$ & $\chi^2$\\
                                     &                    &  type               & (K)              & (K)         &              & (\kms)      &             & (kG) &  (GOF)  \\
\hline
\multirow{3}{*}{GJ~852}              & \multirow{1}{*}{A} & \multirow{1}{*}{M4.0} &3366  & 3282 & -4.30  &  5.30 & 3.5   & 3     & 0.4011 \\
                                     & \multirow{1}{*}{B} & \multirow{1}{*}{M4.5} &3310  & 3281 & -4.30  &  3.91 & 3.5  &  1.5   & 0.1706\\
                                     & \multirow{1}{*}{C} & \multirow{1}{*}{M7.0} &2621  & 2680 & -4.30  &  11.18 &  0.3 & $\sim$& 2.4740\\
\hline                                                                                                                                     
\multirow{2}{*}{GJ~4368 (LHS~4022)}  & \multirow{1}{*}{A} & \multirow{1}{*}{M4.0} &3366&     3315 & -4.89    &     2.71& 3.5   &  0.5 & 0.0634\\
                                     & \multirow{1}{*}{B} & \multirow{1}{*}{M5.5} &3160 &    3187 & -4.89    &     2.49& 3.5   &  0.75&  0.1235\\
\hline                                                                                                                                     
\multirow{2}{*}{LTT~7419}            & \multirow{1}{*}{A} & \multirow{1}{*}{M2.5} &3525&  3535 &-4.79     &   3.41  &  3.5     &     0     & 0.0524\\
                                     & \multirow{1}{*}{B} & \multirow{1}{*}{M7.0} &2621&  2612&  -4.79    &   12.68&  0.3      &    $\sim$ & 2.8217\\
\hline                                                                               
\multirow{2}{*}{GJ~234}              & \multirow{1}{*}{A} & \multirow{1}{*}{M4.5} &3310&  3152& -4.60   &     6.15 & 3.5      &    2.75 &  0.1731\\
                                     & \multirow{1}{*}{B} & \multirow{1}{*}{M7.0} &2621&  2737 &-4.60     &   12.93 &  0.3     &   $\sim$ &0.8577\\
\hline                                                                               
\multirow{2}{*}{LP~717-36}           & \multirow{1}{*}{A} & \multirow{1}{*}{M3.5} &3418&  3517 & -4.37     &    4.07 &   3.5     &   1.75 &0.1172\\
                                     & \multirow{1}{*}{B} & \multirow{1}{*}{M4.0} &3366&  3284 & -4.37    &    12.16 &   3.5    & $\sim$ & 0.6119\\
\hline                                                                               
\multirow{2}{*}{GJ~3322 (LP~476-207)}& \multirow{1}{*}{A} & \multirow{1}{*}{M4.0} &3366&   3418 & -4.40    &    6.94 &   3.5   & 2.75 & 0.1519\\
                                     & \multirow{1}{*}{B} & \multirow{1}{*}{M5.0} &3250&  3197 &-4.40    &    23.80&  3.5       & $\sim$ &0.4048\\
\hline                                                                               
\multirow{2}{*}{GJ~3304}             & \multirow{1}{*}{A} & \multirow{1}{*}{M4.0} &3366&   3336  & -4.10   &     29.26 &   3.5    & $\sim$ & 0.2912\\
                                     & \multirow{1}{*}{B} & \multirow{1}{*}{M4.5} &3310&    3339 & -4.10  &      44.51 &   3.5   & $\sim$  & 0.2621\\
\hline                                                                               
\multirow{2}{*}{GJ~3263 (LHS~1630)}  & \multirow{1}{*}{A} & \multirow{1}{*}{M3.5} &3418&   3417 & -4.80    &     2.10 &   3.5   & 0.56 & 0.0918\\
                                     & \multirow{1}{*}{B} & \multirow{1}{*}{M4.0} &3366&  3356 & -4.80    &     2.53 &   3.5  & 0.5 &0.1147\\
\hline
\multirow{3}{*}{GJ~2005 (LHS~1070)}  & \multirow{1}{*}{A} & \multirow{1}{*}{M5.5} &3160  & 3052 & -4.43  &   6.88& 3.5   & 2 & 0.1613 \\
                                     & \multirow{1}{*}{B} & \multirow{1}{*}{M8.5} &2450  & 2528 & -4.43  &  15.85& 0.3  &  $\sim$   & 5.9142\\
                                     & \multirow{1}{*}{C} & \multirow{1}{*}{M9.0} &2400  & 2547 & -4.43  &  15.04 &  0.3 & $\sim$   & 6.1480\\
\hline
\end{tabular}
\end{center}
\begin{flushleft}
$\teff^{\rm (0)}$ is the temperature which corresponds to the spectral type of the star. \\
No accurate estimates of magnetic field were possible for stars with $\vsini>10$~\kms\ (marked with ``$\sim$'') \\
Last column refer to the chi-square goodness of the fit (GOF) values.
\end{flushleft}
\end{footnotesize}
\label{tab:param-Fe_mean-0-0-1-1}
\end{table*}

\section{Discussion}
\label{sec:discussion}

In our investigation we attempted to derive atmospheric parameters of
seven spectroscopically resolved M-dwarf binary systems and two triple
systems. Using advanced software for the calculation of molecular line
formation in magnetized plasma and high-resolution CRIRES spectra
allowed us to carry out accurate spectroscopic analysis. However, two
main complications exist: 1) the narrow wavelength range of
$\ll\,9920-9970$ observed with CRIRES does not allow an independent
derivation of partially degenerate parameters as $\teff$
and $\abn{Fe}$. We could always find more than one solution. 2)
Unknown van der Waals damping constants of FeH add another uncertainty
to our measurements, especially for slow and moderately rotating stars
where pressure broadening is one of the main line broadening agents.
Thus, we attempted to search for a set of parameters (using chi-square
technique) that provides a satisfactory fit to the data while still
being within our expectations from a physical point of view. For
example, effective temperature is expected to lie within a range
predicted by the spectral type, and iron abundance should not be too
different between components of the same systems and not too different
from the solar one. However, this approach did not always lead to the
best possible agreement between our model and the observed data.
Varying $\teff$ turned out to be an effective way to achieve better
fit quality. In several systems the adopted $\teff$ differs from our
expectations according to spectral types (for example, like M2.5
instead of M3.5 and  M5.5 instead of M4.5 in case of LP~717-36-\ca\ 
and of GJ~234-\ca\ respectively), 
but differences are typically
not larger than one spectral subclass. Such differences can easily be
explained by varying Fe abundances. The distribution of the latter is
usually within 0.3~dex for all systems except GJ~3304 if we assume the
same Fe abundances for components of a given multiple system. 

The possibility of achieving an acceptable fit quality under different
assumptions and finally using quite different values for the atmospheric
structure demonstrates the current difficulty to determine atmospheric
parameters in the spectra of M-type dwarfs. To make progress on this
part, independent information on metallicity or temperature are
required, and improvements to our understanding of the damping
coefficient are needed. Nevertheless, our investigation of different
strategies (Cases 1--4) to fit the spectra shows that the uncertainty
of atmospheric structure does not prevent us from determining the parameters 
we are most interested in this study: rotation velocity and magnetic field strength.


An important result of present spectroscopic investigation is that
such key parameters as $\vsini$ and $\bs$ do not change much between
different fitting cases that we considered. The only exception is
Case~1 as explained in the previous section (fixing $\teff$,
$\abn{Fe}$, and $\fgammaw$ always provided an unreasonable fit). 
Thus, even if there are
systematic uncertainties in estimated $\teff$, $\abn{Fe}$, and/or
$\fgammaw$, this does not significantly influence our results on
$\vsini$ and $\bs$ so that our conclusions remain stable even if based
on imperfect atmospheric descriptions.

\begin{figure*}
\centerline{
\includegraphics[width=0.5\hsize]{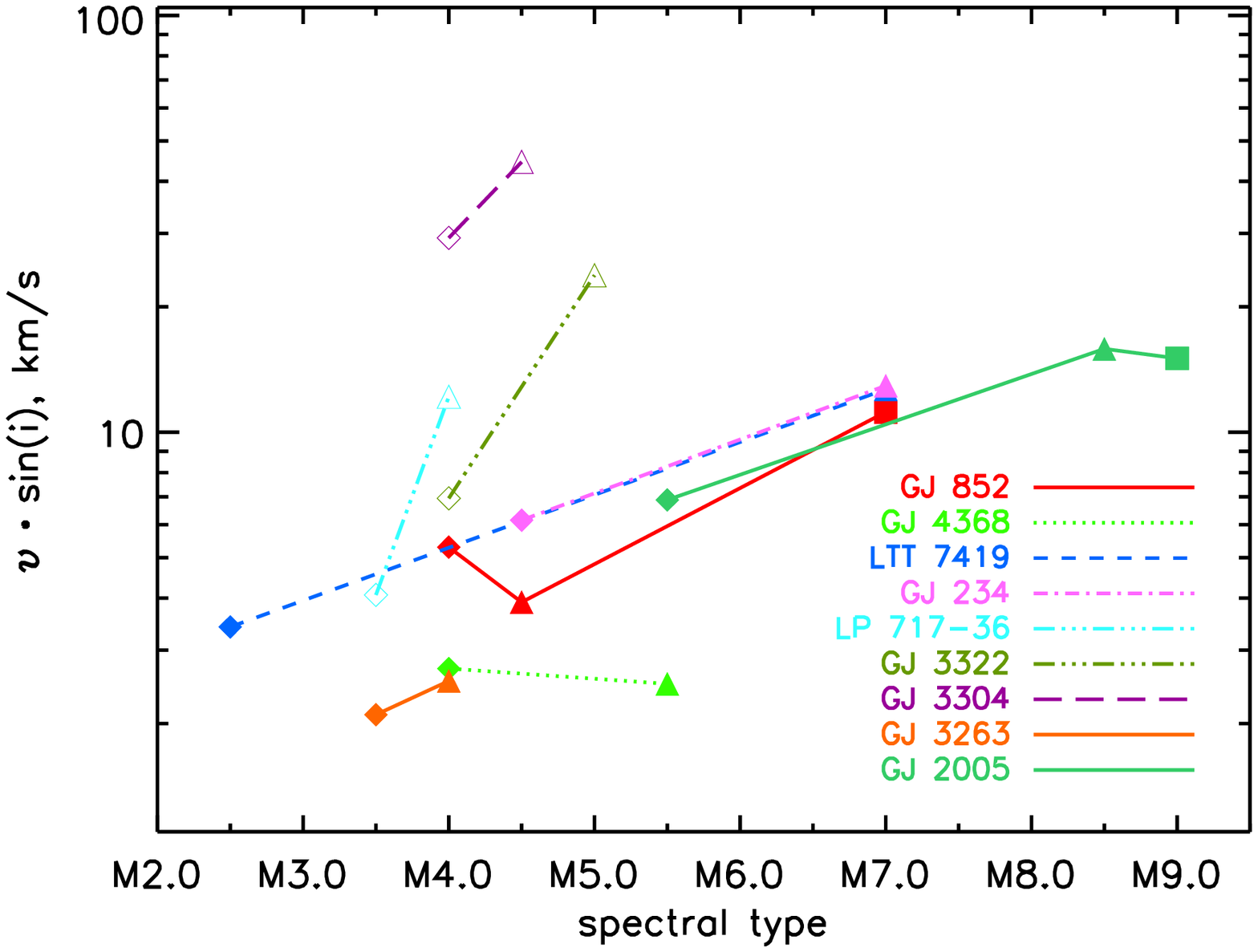}
\includegraphics[width=0.5\hsize]{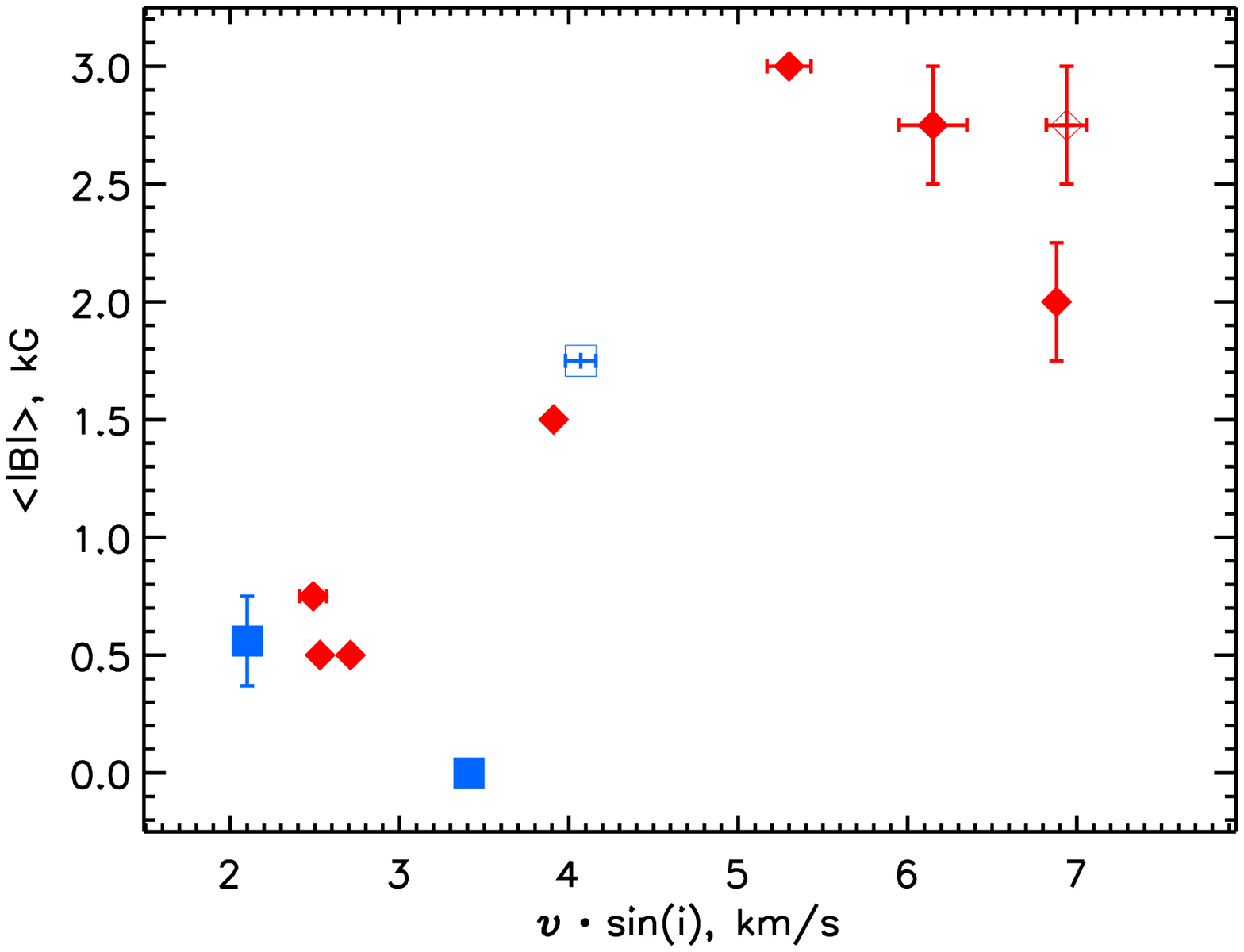}
}
\caption{Graphical representation of main results from Table~\ref{tab:param-Fe_mean-0-0-1-1}.
\textbf{Left plot:} projected rotational velocity $\vsini$ as a function of spectral type. Diamonds~--~primaries,
triangles~--~secondaries, square~--~third components.
\textbf{Right plot:} mean surface magnetic field intensity as a function of $\vsini$ for stars 
with $\vsini<10$~\kms. Blue squares~--~stars of spectral type early than or equal to M4.0; 
red diamonds~--~stars of spectral type later than M4.0. Open symbols indicate young systems.
Error bars correspond to the difference in $\vsini$ and $\bs$ between this case and Case~2.}
\label{fig:res}
\end{figure*}

With our measurements of rotation and magnetism we tried to answer two
main questions:
\begin{enumerate}
\item How efficient is rotational braking in early- and late type components?
\item How is this related to the intensity of the magnetic field?
\end{enumerate}

A first result on spectral type dependent rotational braking in
components of a multiple system was provided by
\citet{2007A&A...471L...5R} showing that the two late-M components in the triple system GJ~2005
rotate significantly faster than the early-M primary. We show the
results of our rotation analysis in the left panel of
Fig.~\ref{fig:res}. In all but one system, the earlier component
rotates at a significantly lower rate than the later secondary. 
We interpret this as clear evidence for rotational
braking being a strong function of spectral type, as expected from
earlier work.
The one exception are the \ca\ and \cb\ components of the triple system GJ~852,
but their spectral types only differ by one subclass and we actually
derive identical temperatures in our fits (even if we assume identical
metallicity). In addition, components \ca\ and \cb\ in GJ~4368 and \cb\ and \cc\ in GJ~2005 
appear to have rotational velocities identical within the uncertainties.

We also have three young systems in our sample: LP~717-36 and GJ~3304 are
likely younger than 300~Myrs \citep{2009ApJ...699..649S}. GJ~3322 is a member
of the $\beta$~Pic moving group and thus probably even younger then 30~Myrs
\citep{2003ApJ...599..342S}. These three systems clearly stand out in Fig.~\ref{fig:res} as the
ones with the highest $\vsini$ and showing the most extreme difference
in rotation rates between their components. The latter is another
indication that braking is less effective at later spectral types, even
more so at young ages.

The reason for the difference in rotational braking is not entirely
clear at this point. Stellar parameters like temperature, mass, and
radius are changing rapidly among mid-M stars, and the change from
partially to fully convective stars adds another complication to the
interpretation because the influence of dynamo on the magnetic geometry 
is unknown. For the slowly rotating objects of our sample, we
provide direct measurements of the average surface magnetic fields. In
fast rotating stars with velocities $\vsini>10$~\kms, magnetic field
estimates are highly uncertain because of strong line blending.
Intensities of the mean surface magnetic fields $\bs$ as a function of
$\vsini$ and spectral type are illustrated in the right panel of
Fig.~\ref{fig:res}. We show only measurements in which Zeeman
broadening was clearly seen and measurable, i.e. $\vsini <
10$\,km\,s$^{-1}$. There is a clear correlation between $\bs$ and
$\vsini$; in our sample of M dwarfs, we find an increase in average
magnetic field strength with $\vsini$. This is an interesting result
because we start to resolve the rising part of the unsaturated
rotation-magnetic field relation \citep{R07}, which was not possible
before because of lower spectral resolving power.

The transition between partial to full convection is believed
to occur around spectral type M3.5.  There is only one system in our
sample with a primary of spectral type earlier than M3.0 and six
with spectral types of M4.0 and earlier. Nevertheless, an effect on
magnetic field generation caused by a change in dynamo mode should not
happen at a sharp threshold but rather take effect smoothly over a
range of stellar masses. For example, all stars later than, say, M5.0
might be expected to have fields much higher than earlier stars at the
same rotation velocity, or the secondaries could have higher fields
even if their rotation velocity does not significantly differ from the
primary's rotation rate. In our sample, we do not find evidence for
any mechanism that requires explanation through the change in dynamo
mode at the boundary to full convection. The differences in average
magnetic field strength can be explained by the influence of rotation
velocity alone. Our sample contains a few stars that are hot enough so that
the influence of a tachocline could still be expected. Their
distribution in velocity shows no difference to the late-M sample that
probably generates magnetism without the presence of a tachocline.

\section{Summary}
\label{sec:summary}
The results of this work can be summarized as follows:
\begin{itemize}
\item
Among nine investigated spectroscopically resolved systems seven clearly have primaries that are
rotating slower than less massive secondaries. In GJ~4368 both \ca\ and \cb\
were found to have comparable rotational velocities, as well as \cb\ and \cc\ in GJ~2005.
Thus, we confirm the existence of spectral type dependent braking in low-mass stars, where the mid- and late
type objects are braked less effectively compared to objects of earlier types.
\item
Three of the systems analysed here are younger than 300~Myrs. In all
three of these system the secondary rotates considerably faster than in
older systems with comparable spectral types, which is fully consistent
with the assumption of differences in the braking efficiency with spectral
type under a general Skumanich-type braking law.
\item
Strong surface magnetic fields were detected in primaries of five systems (GJ~852, GJ~234, LP~717-36, GJ~3322, GJ~2005),
and also in secondary of the triple system GJ~852. No fields could be accurately detected in
components with fast rotation.
\item
For slow and moderately rotating stars, there is a tendency of the mean surface magnetic field to increase with
the rotational velocity $\vsini$.
\item
We find noticeable iron underabundance in such systems as GJ~4368, LTT~7419, and GJ~3263. However,
taking into account other uncertainties (i.e. very fast rotation, $\gammaw$, etc.) this should be considered with
certain caution.
\item
Analysis of the spectra of inactive stars demonstrated that the van der Waals damping constant $\gammaw$
of FeH lines drops significantly in late-type objects compared to its classical value. 
This requires additional and more extensive investigation making use of extended sample of stars and spectra of 
better quality.
\end{itemize}

\section*{Acknowledgments}
This work was supported by the following grants: Deutsche Forschungsgemeinschaft (DFG)
Research Grant RE1664/7-1 to DS and Deutsche Forschungsgemeinschaft under DFG RE 1664/4-1 and NSF
grant AST07-08074 to AS.
OK is a Royal Swedish Academy of Sciences Research Fellow supported by grants
from the Knut and Alice Wallenberg Foundation and the Swedish Research Council.
We also acknowledge the use of electronic databases (VALD, SIMBAD, NASA's ADS).
This research has made use of the Molecular Zeeman Library (Leroy, 2004).

\bsp
\label{lastpage}

\end{document}